# Price and Payoff Autocorrelations in a Multi-period Consumption-Based Asset Pricing Model

Victor Olkhov

Independent, Moscow, Russia

victor.olkhov@gmail.com

ORCID: 0000-0003-0944-5113

March 20, 2024.

## Abstract

This paper highlights the hidden dependence of the basic pricing equation of a multi-period consumption-based asset pricing model on price and payoff autocorrelations. We obtain the approximations of the basic pricing equation that describe the mean price "to-day," mean payoff "next-day," price and payoff volatilities, and price and payoff autocorrelations. The deep conjunction of the consumption-based model with other versions of asset pricing, such as ICAPM, APM, etc. (Cochrane, 2001), emphasizes that our results are valid for other pricing models.

This research received no support, specific grant or financial assistance from funding agencies in the public, commercial or nonprofit sectors. We welcome valuable offers of grants, support and positions.

## 1. Introduction

This paper, for the consumption-based pricing model, emphasizes the dependence of the basic pricing equation on price and payoff autocorrelations.

Investors are looking for "signs" that highlight the price movements up or down. These efforts support the development of pricing models that could "predict" price variations and lead to the description of price autocorrelations. Studies on market price correlations are part of a more general problem of estimating correlations between different economic and financial variables (Kendall and Hill, 1953; Fama, 1965; Campbell, Grossman, and Wang, 1992; Liu et al., 1997; Plerou et al., 2000; Quinn and Voth, 2008; Diebold and Strasser, 2010; Lind and Ramondo, 2018). Any attempt to review the current state of economic correlation studies requires separate and deep research. This paper is not an introduction for beginners, and we assume that our readers are familiar with conventional asset pricing studies (Sharpe, 1964; Merton, 1973; Friedman, 1990; Campbell, 2000; Cochrane, 2001; Cochrane and Culp, 2003; Barillas and Shanken, 2018). We propose that readers have experience with the basics of probability theory, statistical moments, etc.

The description of the price autocorrelation is an important part of the random market price studies. We consider the multi-period consumption-based pricing model and show that the usual assumptions on the utility function lead to the expressions of price and payoff autocorrelation. The frame of the consumption-based model (Cochrane, 2001) could generate other versions of asset pricing, like Intertemporal CAMP (ICAPM), Arbitrage pricing theory (APT), etc. Thus, our results can also be obtained for other asset pricing models.

The main contribution of this paper is the derivation of the approximations of the basic pricing equations for a multi-period model that describe the mean price today at time $t$, the mean payoff "next day" at time $T$, and price and payoff volatilities at times $t$ and $T$. We highlight the dependence of the basic pricing equations on price autocorrelation and payoff autocorrelation.

## 2. Covert issues of the asset pricing

We consider common consumption-based asset pricing and follow a well-known manual (Cochrane, 2001) as the main source for any details. Cochrane's study presents a clear and complete description of the pricing models, describes numerous cases, and demonstrates the unity of most variations of asset pricing. Thus, we believe that our new results can be derived from other variations of asset pricing. We refer to Cochrane (2001) as the basic source and show that the initial assumptions on the price averaging procedure and the form of the utility function hide certain issues that generate important pricing relations. We present simple modifications of the usual asset averaging procedure and investor's utility function and derive extensions of the basic equation. That causes modification of the "main statement" of the asset pricing: "Price equals expected discounted payoff" and presents assessments of price autocorrelation.

Let us start with the utility $u(c_t)$ at current day $t$ and the utility $u(c_T)$ "next day" $T$ and keep almost all notations (Cochrane, 2001):

$$U(c_t;\ c_T) = u(c_t) + \beta E[u(c_T)] \tag{2.1}$$

$$c_t = e_t - p\xi \quad ; \quad c_T = e_T + x_T\,\xi \quad ; \quad x_T = p_T + d_T \tag{2.2}$$

$E[..]$ denotes math expectation at day $T$ under the information available at day $t$, and $\beta$ is a subjective discount factor. $c_t$ and $c_T$ denote consumptions at days $t$ and $T$; $p_t$ and $p_T$ denote asset prices at $t$ and $T$; $d_T$ and $x_T$ denote dividends and payoffs at day $T$. $e_t$ and $e_T$ denote consumption at days $t$ and at $T$ without investments. $\xi$ denotes the amount of assets the investor purchases at day $t$ and sells at day $T$. The consumption-based pricing model determines the basic pricing equation (2.4) via the condition of the maximum utility (2.1) by the amount of assets $\xi$:

$$\max_{\xi} U(c_t;\ c_{t+1}) \leftrightarrow \frac{\partial}{\partial \xi} U(c_t;\ c_{t+1}) = 0 \tag{2.3}$$

$$p = \beta E\left[\frac{u'(c_T)}{u'(c_t)}\, x\right] = E[mx] \tag{2.4}$$

$$m = \beta \frac{u'(c_T)}{u'(c_t)} \quad ; \quad u'(c) = \frac{d}{dc} u(c) \tag{2.5}$$

The basic pricing equation (2.4) is the origin of the popular statement: "Price equals expected discounted payoff" (Cochrane, 2001). Let us reconsider (2.1-2.5), taking into account the economic meaning of mathematical expectation $E[..]$.

Indeed, any market trade data is presented as a time series, and thus, any averaging procedure of these time series should aggregate a certain number of their terms during a certain time interval $\Delta$. Let the market price $p(t_i)$ time series be determined at times $t_i$ with a time shift ε :

$$t_i = \varepsilon \cdot i \quad ; \quad i = 0, 1, 2, \dots \tag{2.6}$$

The time scale $\varepsilon$ between two market trades determines the market-based division of the time axis. Aggregation or averaging of time series during the interval $\Delta$ replaces the initial time axis divisions multiple of $\varepsilon$ by the time divisions multiple of the interval $\Delta$. Such a change of the time axis divisions cannot be performed only on "the next day" $T$, but it should also be performed today, on day $t$. It seems reasonable that the economic problems should be described on the time axis with unified divisions along the entire time axis. Thus, any averaging procedure of economic time series during the interval *Δ* that is performed by mathematical expectation *E[..]* at day *T* should be complemented by the similar averaging procedure during the same interval *Δ* today, at day *t*. That replaces the investor's utility function (2.1) by:

$$U(c_t;\ c_T) = E[u(c_t)] + \beta E[u(c_T)] \tag{2.7}$$

In (2.7), $E[u(c_t)]$ denotes mathematical expectation for the day *t,* and $E[u(c_T)]$ denotes mathematical expectation for the day *T.* Both mathematical expectations are performed during the same averaging interval *Δ.* Averaging during the same interval *Δ* today, at day *t,* and "next day" at *T* establishes the same divisions of the time axis as a multiple of *Δ*.

If the amount of assets *ξ* delivers the maximum to the investor's utility (2.7), then (2.3; 2.4) are replaced by the modified basic equation (2.8):

$$E[u'(c_t)p] = \beta E\ [u'(c_T)x] \tag{2.8}$$

The symbol *E[..]* in the left side of (2.8) denotes mathematical expectation at day *t,* and *E[..]* in the right side (2.8) denotes mathematical expectation at day *T* under the information available at day *t.* The direct assessment of mathematical expectations in (2.8) is a difficult problem. To simplify it, let us derive an approximation of (2.8) using a simple Taylor series.

## 3. Approximation of the basic pricing equation

To derive an approximation of the basic equation (2.8), we present the utility functions during the averaging interval *Δ* at day *t* and at day *T* using Taylor series by the variations of price at day *t* and by the variations of payoff at day *T* (Olkhov, 2021). Indeed, averaging *E[..]* during *Δ* assumes that price *p* at day *t* and payoff *x* at day *T* during *Δ* can be presented as:

$$p\ =\ p_0 + \delta p \quad ; \quad x = x_0 + \delta x \tag{3.1}$$

Here $p_0$ is a mean price during the averaging interval *Δ* at day *t,* and $\delta p$ is price variation near the mean $p_0$ during *Δ*. We use similar notations for the mean payoff $x_0$ and the payoff variations $\delta x$ during *Δ* at day *T*. The relations (3.2) determine the mean price $p_0$, mean payoff $x_0$, price volatility $\sigma_p^2(t)$ at day *t,* and payoff volatility $\sigma_x^2(T)$ at day *T*:

$$p_0 = E[p]\ ;\ x_0 = E[x]\ ;\ E[\delta p] = E[\delta x] = 0\ ;\ \sigma_p^2(t) = E[\delta^2 p]\ ;\ \sigma_x^2(T) = E[\delta^2 x] \tag{3.2}$$

The relations (2.2; 3.1; 3.2) allow us to present the Taylor series for the utility functions (2.8). Here we consider only the linear expansion of the Taylor series:

$$u'(c_t) = u'\left(c_{t;0}\right) - u''\left(c_{t;0}\right)\,\xi\,\delta p \quad ; \quad c_{t;0} = e_t - p_0\xi \tag{3.3}$$

$$u'(c_T) = u'\left(c_{T;0}\right) + u''\left(c_{T;0}\right)\,\xi\,\delta x \quad ; \quad c_{T;0} = e_T + x_0\,\xi \tag{3.4}$$

The substitution of the linear Taylor series (3.3; 3.4) of the utility functions (2.7) into (2.8) and (3.2) gives the linear approximation of the basic pricing equation (2.8) as:

$$p_0(t) = \beta \frac{u'(c_{T;0})}{u'(c_{t;0})} x_0(T) + \beta\xi \frac{u''(c_{T;0})}{u'(c_{t;0})} \sigma_x^2(T) + \xi \frac{u''(c_{t;0})}{u'(c_{t;0})} \sigma_p^2(t) \qquad (3.5)$$

The approximation of the basic equation (3.5) establishes the direct dependence of the mean price $p_0(t)$ at day $t$ on the discounted mean payoff $x_0(T)$ at day $T$. However, the approximation (3.5) also determines the dependence of the mean price $p_0$ at day $t$ on the price volatility $\sigma_p^2(t)$ at day $t$ and on the payoff volatility $\sigma_x^2(T)$ at day $T$. Taking into account that the second derivative of the utility always should be negative, obtain the obvious condition that the growth of the price volatility $\sigma_p^2(t)$ or payoff volatility $\sigma_x^2(T)$ should be lower than the mean price $p_0$ at day $t$.

It can be noted that (3.5) presets the direct linear dependence of the mean price $p_0$ at day $t$ on the amount of assets $\xi$. However, the dependence on $\xi$ is also hidden in the form of the derivatives of the utility functions (3.3; 3.4) and in the basic pricing equation (2.4) by the dependence of consumption $c_t$ at day $t$ and consumption $c_T$ at day $T$ (2.2) on the amount of assets $\xi$. It is obvious that the discount factor $m$ (2.5) (Cochrane, 2001) has a hidden dependence on the amount of assets $\xi$ that delivers the maximum to the investor's utility (1.1). The complexity of that dependence and the complexity of the assessment of the mathematical expectation in (2.5) that takes into account dependence on $\xi$ result in omitting these relations and consideration of the discount factor $m$ (2.5) as "given." Actually, even the definition of the amount of $\xi$ that delivers the maximum to the investor's utility (1.1) is a tough problem that requires separate calculations Olkhov (2021).

Now we consider modifications to the utility (2.7) that help assess price autocorrelation.

## 4. Price autocorrelation

Economic considerations that justify the investor's utility function in the form of (2.1) or (2.7) are rather simple. The investor chooses between the consumption $e_t$ at day $t$ and the consumption $e_T$ at day $T$ (2.2). The investor's utilities (2.1; 2.7) and the basic equations (2.4; 2.8) model the case with a single purchase and a single sale of assets. However, the investor can make a decision to perform two purchases of assets during some time interval $l$ at day $t$ and sell all assets at day $T$. How can one model that case using the investor's utility function (2.1) or (2.7)?

We model two purchases and a single sale of assets by two utilities and two basic equations. The model of the first purchase of the assets at time $t_1$ and the sale at day $T$ coincides (4.2) with the above utility (2.7) and consumption (2.2):

$$c_{t_1} = e_{t_1} - p(t_1)\xi(t_1) \quad ; \quad c_T = e_T + x_1\,\xi(t_1); \;\; x_1 = x_{1;0} + \delta x_1 \qquad (4.1)$$

$$E[u'(c_{t_1})p(t_1)] = \beta E\,[u'(c_T)x_1] \qquad (4.2)$$

In (4.1; 4.2), $x_{1;0}$ and $\delta x_1$ denote the assessments of the mean payoff and payoff variations under the information available at time $t_1$.

We describe the second purchase of the assets at time $t_2$ and the sale at the same day $T$ by utility (2.7) and consumption (4.2). At time $t_2$, the forecast of the mean payoff and payoff variations can be different from the forecast made at time $t_1$. Thus, at time $t_2$, we model consumption as:

$$c_{t_2} = e_{t_1} - p(t_1)\xi(t_1) - p(t_2)\xi(t_2) \;\; ; \;\; c_T = e_T + x_2[\,\xi(t_1) + \xi(t_2)] \;\; ; \;\; t_2 = t_1 + l \qquad (4.3)$$

The relations (4.3) model the assumption that the investor does not change his consumption since time $t_1$ during the interval $l$ that defines the time $t_2$ of the second purchase. The relations (4.3) don't change the form of the basic equation (2.8), and (4.2) is determined by the maximum condition of the utility (2.7) by the amount of assets $\xi(t_2)$:

$$E[u'(c_{t_2})p(t_2)] = \beta E\;[u'(c_T)x_2] \qquad (4.4)$$

The basic equations (4.2) and (4.4) describe the maximum of the investor's utility (2.7) for the first and second purchases and are similar to the basic equation (2.8) of a single purchase. It seems that nothing has changed. However, linear Taylor series approximations show that (4.3) and (4.4) give assessments of the price autocorrelation. To show that we present the price and payoff as:

$$p(t_1) = p_0(t_1) + \delta p_1 \quad ; \quad p(t_2) = p_0(t_2) + \delta p_2 \;\; ; \;\; x_2 = x_{2;0} + \delta x_2 \qquad (4.5)$$

The linear Taylor series approximations of the utilities in (4.4) due to (4.3; 4.5) give:

$$u'(c_{t_2}) = u'(c_{t_2;0}) - u''(c_{t_2;0})[\xi(t_1)\delta p_1 + \xi(t_2)\,\delta p_2] \qquad (4.6)$$

$$u'(c_T) = u'(c_{T;0}) + u''(c_{T;0})\;[\xi(t_1) + \xi(t_2)]\;\;\delta x_2 \qquad (4.7)$$

$$c_{t_2;0} = e_{t_1} - p_0(t_1)\xi(t_1) - p_0(t_2)\xi(t_2) \quad ; \quad c_{T;0} = e_T + x_{2;0}[\,\xi(t_1) + \xi(t_2)] \qquad (4.8)$$

$$p_0(t_1) = E[p(t_1)] \;;\; p_0(t_2) = E[p(t_2)] \,;\; \sigma_p^2(t_1) = E\left[\delta^2 p_1\right] \quad ; \quad \sigma_p^2(t_2) = E\left[\delta^2 p_2\right] \qquad (4.9)$$

$$\sigma_{x_2}^2(T) = E\left[\delta^2 x_2\right] \;\; ; \;\; corr\{p(t_1)p(t_2)\} \;=\; E[\delta p_1 \delta p_2] \;\; ; \;\; E[\delta p_1] = E[\delta p_2] = 0 \qquad (4.10)$$

The term $corr\{p(t_1)p(t_2)\}$ in (4.10) describes the price autocorrelation at times $t_1$ and $t_2$. Taking into account (3.2; 4.9; 4.10), one obtains that the approximation (4.11) of the basic pricing equation that describes the first purchase at time $t_1$ is similar to (3.5):

$$p_0(t_1) = \beta\frac{u'(c_{T;0})}{u'(c_{t_1;0})}x_{1;0} + \beta\frac{u''(c_{T;0})}{u'(c_{t_1;0})}\xi(t_1)\sigma_{x_1}^2(T) + \frac{u''(c_{t_1;0})}{u'(c_{t_1;0})}\xi(t_1)\sigma_p^2(t_1) \qquad (4.11)$$

The basic equation (4.11) describes the dependence of the mean price $p_0(t_1)$ on mean payoff $x_{1;0}$ at time $T$, on the amount of assets $\xi(t_1)$ that delivers the maximum to the utility (2.7), and on the volatility $\sigma_p^2(t_1)$ of price at time $t_1$ and on the volatility of payoff $\sigma_{x1}^2(T)$ at time $T$. Now let us substitute (3.2; 4.5 - 4.10) in (4.4), and in the linear approximation by the Taylor series of (4.4), obtain the basic equation (4.12) that introduces the dependence of the mean price $p_0(t_2)$ on price autocorrelation $corr\{p(t_1)p(t_2)\}$:

$$p_0(t_2) = \beta\frac{u'(c_{T;0})}{u'(c_{t_2;0})}x_{2;0} + \beta\frac{u''(c_{T;0})}{u'(c_{t_2;0})}[\,\xi(t_1) + \xi(t_2)]\sigma_{x_2}^2(T) +$$

$$+\frac{u''(c_{t_2;0})}{u'(c_{t_2;0})}\,[\,\xi(t_1)corr\{p(t_1)p(t_2)\} + \xi(t_2)\sigma_p^2(t_2)] \qquad (4.12)$$

We highlight that the mean payoff $x_{2;0}$ and the payoff volatility $\sigma_{x2}{}^2(T)$ at time $T$ in the basic equation (4.12) for the second purchase at time $t_2$ can be different from the values that describe the basic equation (4.11) for the first purchase at time $t_1$ because they are obtained under the information at different times $t_1$ and $t_2$.

If the time interval $l$ between the first purchase at time $t_1$ and the second purchase at time $t_2 = t_1+l$ tends to zero $l \rightarrow 0$ then the amount of assets of the second purchase $\xi(t_2) \rightarrow 0$ also tends to zero. Indeed, the first purchase of $\xi(t_1)$ assets delivers max to utility (2.7), and hence no more assets are required during the small time interval $l= t_2 - t_1$. In that case, for $l \rightarrow 0$, $\xi(t_2) \rightarrow 0$ one can neglect $\xi(t_2)$ to compare with $\xi(t_1)$ and equation (4.12) for $t_2 \rightarrow t_1$ takes the form:

$$p_0(t_1) = \beta\frac{u'(c_{T;0})}{u'(c_{t_1;0})}x_{1;0} + \beta\frac{u''(c_{T;0})}{u'(c_{t_1;0})}\;\xi(t_1)\sigma_{x_1}^2(T) + \frac{u''(c_{t_1;0})}{u'(c_{t_1;0})}\;\xi(t_1)\sigma_p^2(t_1) \qquad (4.13)$$

As required, due to (4.11), the price autocorrelation in (4.13) matches $corr\{p(t_1)p(t_1)\}= \sigma_p{}^2(t_1)$. Readers can easily derive the step-by-step modifications of the basic pricing equation that describe the case when the investor performs $k$ successive purchases of assets at moments $t_1,\ldots t_k$ and then sells all assets at moment $T$. In this case, the basic pricing equation will depend on price autocorrelations $corr\{p(t_i)p(t_j)\}$, $i, j \leq k$.

## 5. Payoff autocorrelation

A consumption-based asset pricing model that uses the investor's utility (2.8) can describe the payoff autocorrelation. To show that, let us consider the case when the investor makes two successive purchases of assets at times $t_1$ and $t_2$ and then performs two successive sales of assets at times $T_1$ and $T_2$. We model the first purchase and first sale of assets similar to (4.1; 4.2):

$$c_{t_1} = e_{t_1} - p(t_1)\xi(t_1) \quad ; \quad c_{T_1} = c_{T_1} + x_1(T_1)\xi(t_1) \;\; ; \;\; x_{11}(T_1) = x_{11;0}(T_1) + \delta x_{11} \quad (5.1)$$

At time $t_1$ at price $p(t_1)$, the investor purchases the amount $\xi(t_1)$ and sells the amount $\xi(t_1)$ at time $T_1$. We denote $x_{11}(T_1)$ as the payoff, $x_{11;0}(T_1)$ as the mean payoff, and $\delta x_{11}$ as the payoff variations predicted at time $T_1$ under the information available at time $t_1$. The investor's utility at time $t_1$ takes the form:

$$U(c_{t_1};\, c_{T_1}) = E[u(c_{t_1})] + \beta E[u(c_{T_1})] \qquad (5.2)$$

The amount $\xi(t_1)$ of assets delivers the maximum to the investor's utility (5.2) and determines the basic pricing equation:

$$E[u'(c_{t_1})p(t_1)] = \beta E\left[u'(c_{T_1})x_{11}(T_1)\right] \tag{5.3}$$

Then, at time $t_2 = t_1 + l$, the investor purchases the amount $\xi(t_2)$ of assets and sells these assets at time $T_2 = T_1 + L$ with the payoff $x_2(T_2)$ that was predicted under the information available at time $t_2$. We model that case as follows:

$$c_{t_2} = e_{t_1} - p(t_1)\xi(t_1) - p(t_2)\xi(t_2) \quad ; \quad c_{T_2} = c_{T_1} + x_{12}(T_1)\xi(t_1) + x_2(T_2)\xi(t_2) \tag{5.4}$$

We denote $x_{12}(T_1)$ as the payoff, $x_{12;0}(T_1)$ as the mean payoff, and $\delta x_{12}$ as the payoff variations at time $T_1$ predicted under the information available at time $t_2$. As $x_2(T_2)$, $x_{2;0}(T_2)$, and $\delta x_2$, we denote the payoff, mean payoff, and payoff variations at time $T_2$ predicted under information available at time $t_2$.

$$x_{12}(T_1) = x_{12;0}(T_1) + \delta x_{12} \quad ; \quad x_2(T_1) = x_{2;0}(T_1) + \delta x_2 \tag{5.5}$$

The investor's utility at time $t_2$ takes the form:

$$U(c_{t_2};\, c_{T_2}) = E[u(c_{t_2})] + \beta E[u(c_{T_2})] \tag{5.6}$$

The relations (5.4; 5.5) and the assumption that $\xi(t_2)$ at time $t_2$ delivers the maximum to the investor's utility (5.6) give the basic equation (5.7):

$$E[u'(c_{t_2})p(t_2)] = \beta E\left[u'(c_{T_2})x_2(T_2)\right] \tag{5.7}$$

The linear Taylor series of the price and payoff variations of the utility functions, similar to (4.6; 4.7), gives the approximations of the basic equations (5.2; 5.5):

$$p_0(t_1) = \beta \frac{u'(c_{T1;0})}{u'(c_{t_1;0})} x_{11;0}(T_1) + \beta \frac{u''(c_{T1;0})}{u'(c_{t_1;0})} \xi(t_1)\sigma^2_{x_{11}}(T_1) + \frac{u''(c_{t_1;0})}{u'(c_{t_1;0})} \xi(t_1)\sigma^2_p(t_1) \tag{5.6}$$

$$\sigma^2_{x_{11}}(T_1) = E\left[\delta^2 x_{11}\right] \quad ; \quad \sigma^2_{x_2}(T_1) = E\left[\delta^2 x_2\right] \tag{5.7}$$

$$p_0(t_2) = \beta \frac{u'(c_{T2;0})}{u'(c_{t_2;0})} x_{12;0} + \beta \frac{u''(c_{T2;0})}{u'(c_{t_2;0})} \left[\, \xi(t_1) corr\{x(T_1)x(T_2)\} + \xi(t_2)\, \sigma^2_{x_2}(T_2)\right] +$$

$$+ \frac{u''(c_{t_2;0})}{u'(c_{t_2;0})} \left[\, \xi(t_1) corr\{p(t_1)p(t_2)\} + \xi(t_2)\sigma^2_p(t_2)\right] \tag{5.8}$$

The payoff autocorrelation $corr\{x(T_1)x(T_2)\}$ is determined by the mathematical expectation (5.9) of the payoff variations $\delta x_{12}$ at time $T_1$ and $\delta x_2$ at time $T_2$ predicted under the information available at time $t_2$.

$$corr\{x(T_1)x(T_2)\} = E[\delta x_{12}(T_1)\delta x_2(T_2)] \tag{5.9}$$

The price autocorrelation $corr\{p(t_1)p(t_2)\}$ is determined by (4.10).

Readers can extend the above results and derive the step-by-step modifications of the basic pricing equation that describe the case when the investor performs $k$ successive purchases of assets at times $t_1, \ldots t_k$ and then k successive sales of assets at times $T_1, \ldots T_k$. In this case, the basic pricing equation will depend on price autocorrelations $corr\{p(t_i)p(t_j)\}$ and payoff autocorrelations $corr\{x(T_i)x(T_j)\}$, $i, j \leq k$.

## 6. Conclusion

This paper emphasizes the dependence of the basic pricing equations (4.12) and (5.8) on price (4.10) and payoff autocorrelation (5.9). If it is acknowledged that the multi-period consumption-based pricing model could give rise to alternative pricing models like Intertemporal CAMP (ICAPM), Arbitrage pricing theory (APT) etc., then the above results valid for other pricing models. One should consider the averaging interval $\Delta$ as the starting point of any asset pricing model and as a necessary tool for the averaging or smoothing of time series. The use of Taylor series expansions during the averaging interval helps to consider two or more serial trades with assets and to derive the above results using different versions of pricing models.

## References

Barillas, F. and J. Shanken, (2018). Comparing Asset Pricing Models. J.Finance, 73 (2), 715-754

Campbell, J.Y., Grossman, S.J. and J.Wang, (1992). Trading Volume and Serial Correlation in Stock Return. NBER WP 4193, Cambridge, MA., 1-45

Campbell, J.Y. (2000). Asset Pricing At The Millennium. NBER, WP 7589, Cambridge, 1-75

Cochrane, J.H. (2001). Asset Pricing. Princeton Univ. Press, Princeton, US

Cochrane, J.H., Culp, C.L. (2003). Equilibrium Asset Pricing and Discount Factors: Overview and Implications for Derivatives Valuation and Risk Management. In: Modern Risk Management. A History, Ed. S.Jenkins, 57-92

Diebold, F.X. and G. Strasser, (2010). On The Correlation Structure Of Microstructure Noise: A Financial Economic Approach, NBER, WP 16469, Cambridge, MA, 1-65

Fama, E.F. (1965). The Behavior of Stock-Market Prices. J. Business, 38 (1), 34-105

Friedman, D.D. (1990). Price Theory: An Intermediate Text. South-Western Pub. Co., US

Kendall, M.G and A.B. Hill, (1953). The Analysis of Economic Time-Series-Part I: Prices. Jour. Royal Statistical Soc., Series A, 116 (1), 11-34

Lind, N. and N. Ramondo, (2018). Trade With Correlation, NBER, WP 24380, Cambridge, MA, 1-64

Liu,Y., Cizeau, P., Meyer, M., Peng, C-K. and H. E. Stanley, (1997). Correlations in Economic Time Series. Physica A, 245, 437-440

Merton, R.C. (1973). An Intertemporal Capital Asset Pricing Model, Econometrica, 41, (5), 867-887

Olkhov, V. (2021). Three Remarks On Asset Pricing. SSRN WP3852261, 1-20

Plerou, V., Gopikrishnan, P., Rosenow, B., Amaral, L.A.N. and H. Stanley, (2000). Econophysics: Financial time series from a statistical physics point of view, Physica A, 279 443-456

Quinn, D.P. and H-J. Voth, (2008). A Century of Global Equity Market Correlations. American Economic Review, 98 (2), 535–540

Sharpe, W.F., (1964). Capital Asset Prices: A Theory of Market Equilibrium under Conditions of Risk, Jour. Finance, 19 (3), 425-442